\documentclass[twocolumn,preprintnumbers,amsmath,amssymb]{revtex4-1}
\usepackage{graphicx}
\usepackage{dcolumn}
\usepackage{amsmath,amsbsy,amssymb,scalefnt}
\usepackage{bm}
\usepackage{epstopdf}
\usepackage{amsfonts}
\usepackage{textcomp}
\usepackage{mathtools}
\usepackage{hyperref}
\hypersetup{backref=true,
 pdfnewwindow=true, colorlinks=true,
 linkcolor=blue, anchorcolor=blue,
 citecolor=blue, filecolor=blue,
 menucolor=blue, urlcolor=blue}

\begin{document}


\title{A tight-binding investigation of biaxial strain induced topological phase transition in GeCH$_3$}

\author{Mohsen Rezaei$^1$}
\thanks{These two authors contributed equally.}
\author{Esmaeil Taghizadeh Sisakht$^{1,2}$}
\thanks{These two authors contributed equally.}
\author{Farhad Fazileh$^1$}
\email{fazileh@cc.iut.ac.ir}
\author{Zahra Aslani$^1$}
\author{F. M. Peeters$^2$}
\affiliation{$^1$Department of Physics, Isfahan University of Technology, Isfahan 84156-83111, Iran\\
$^2$Department of Physics, University of Antwerp, Groenenborgerlaan 171, B-2020 Antwerpen, Belgium}

\date{\today}

\begin{abstract}
 We propose a tight-binding (TB) model, that  includes spin-orbit coupling (SOC), 
 to describe the electronic properties 
 of methyl-substituted germanane (GeCH$_3$). 
 This model gives an electronic spectrum in agreement with first principle results
 close to the Fermi level.
  Using the $\mathbb{Z}_2$  formalism,
 we show that a topological phase transition from a normal insulator (NI) to a
 quantum spin Hall (QSH) phase occurs at 11.6\% biaxial tensile strain. 
 The sensitivity of the electronic properties of this system on strain, in particular 
 its transition to the topological insulating phase, makes it very attractive 
 for applications in strain sensors and other microelectronic applications.   
\end{abstract}

\maketitle


\section{\label{introduction}Introduction}

Topological insulators (TIs) are a new state of matter that have attracted 
a lot of interest within the condensed matter physics community~\cite{kane,hasan,qi,moor,fu,bernevig}.
It is now well established that TIs are promising candidates for future  advanced electronic 
devices. They possess a bulk insulating gap and conducting edge states. The edge states are protected 
by time-reversal symmetry (TRS) against backscattering and this property makes them robust against 
disorder and nonmagnetic defects. Consequently, the edge channels normally possess  very high carrier mobility. 

Among TI materials two-dimensional (2D) van der Waals systems have attracted a lot of attention during 
the past decade~\cite{mannix2017}. The interest in these systems  originates from the discovery of graphene,
which has a very high carrier mobility (200 000 cm$^2$/(V s)), thermal conductivity, and mechanical
strength~\cite{neto2009,katsnelson2012}; however, its zero electronic band gap has severely limited its 
applicability in electronic devices. Also, the proposal for the existence of a topological insulating phase in graphene by Kane
and Mele was shown to be unrealistic, because of its extremely small SOC strength~\cite{yao2007,huertas2006}. 
Hence, extensive efforts have been devoted to open a band gap and increase the effective SOC
in graphene or find other 2D systems with favorable SOC, carrier mobility, and appropriate band gap. 

Other 2D materials such as single- or few-layer transition metal dichalcogenides (TMDs), boron nitride,
silicene, germanene, phosphorene, stanene and MXene, have been extensively explored~\cite{ren2016,mannix2017}. 

Another important issue for applications in electronic industry is the compatibility of the material
with current silicon-based electronic technology. Therefore, the group IV elements with honeycomb
structure are more favorable for this purpose. 

One method for tuning the electronic band structure of 2D systems is the use of surface functionalization.
Functionalization of graphene with hydrogen, the so-called hydrogen-terminated graphene or graphane,
opens a sizeable band gap, but its carrier mobility decreases dramatically to 10 cm$^2$/(Vs)~\cite{elias2009control}.
Silicene and germanene the other analogues of graphene have also attracted much attention.
However, the small band gap of these systems and  mobility issues have limited their application for electronics.
Functionalized germanene provide enhanced stability and tunable properties~\cite{jiang2014improving}.
Compared with bulk Ge, surface functionalized germanene possess a direct and large band gap depending
on the surface ligand. These materials can be synthesized via the topotactic deintercalation of
layered Zintl phase precursors~\cite{jiang2014improving,jiang2014covalently}. In contrast to TMDs, 
the weaker interlayer interaction allows for direct band gap single layer properties such as
 strong photoluminescence that are readily present without the need to exfoliate down to a single layer.
Bianco et al.~\cite{bianco2013stability}  produced experimentally hydrogen-terminated 
germanene, GeH (also called, germanane). Recently the new material GeCH$_3$
was synthesized~\cite{jiang2014improving}, that exhibit an enhanced thermal stability. GeCH$_3$ 
is thermally stable up to $250\,^{\circ}\mathrm{C}$ which compares to $75\,^{\circ}\mathrm{C}$ for GeH.
The electronic structure of GeCH$_3$ has been shown to be very sensitive to strain, which makes it very attractive
for strain sensor applications~\cite{ma2014strain,jing2015high,ma2016band}. It has also a high carrier mobility
and pronounced light absorption which makes it attractive for light harvesting applications~\cite{jing2015high,ma2016band}.

At present there exist already a few first-principle studies of GeCH$_3$  that also include the effect of
SOC~\cite{jiang2014improving,ma2014strain,jing2015high,ma2016band}.
To fully understand the physics behind the electronic band structure close to the Fermi level, 
we propose a TB model.
Our TB model is fitted to the density functional theory (DFT) results both for the case with and
without SOC. In the next part of this work we applied biaxial
tensile strain to examine the effect of strain on the electronic properties of this system and 
compare our results with DFT calculations. The possibility of a topological 
phase transition in GeCH$_3$ under  biaxial tensile strain is also examined. Our finding that there is a transition
to the QSH phase is further corroborated by the fact that we find TRS protected edge 
states in nanoribbons made out of GeCH$_3$.

This paper is organized as follows. In Sec.~II, we introduce the crystal structure and
lattice constants of  monolayer GeCH$_3$. Our TB model with and without SOC is introduced in 
Sec.~III, 
and the effect of strain on the electronic properties of  monolayer GeCH$_3$ 
is examined. In Sec.~IV, using the $\mathbb{Z}_2$  formalism we demonstrate 
the existence of a topological phase transition in the electronic properties
of  monolayer GeCH$_3$ when biaxial tensile strain is applied.
The paper is summarized in Sec.V.
 
\section{\label{structure}lattice  structure of monolayer G\lowercase{e}CH$_3$ }

\begin{figure}
\centering
\includegraphics[width=.46\textwidth]{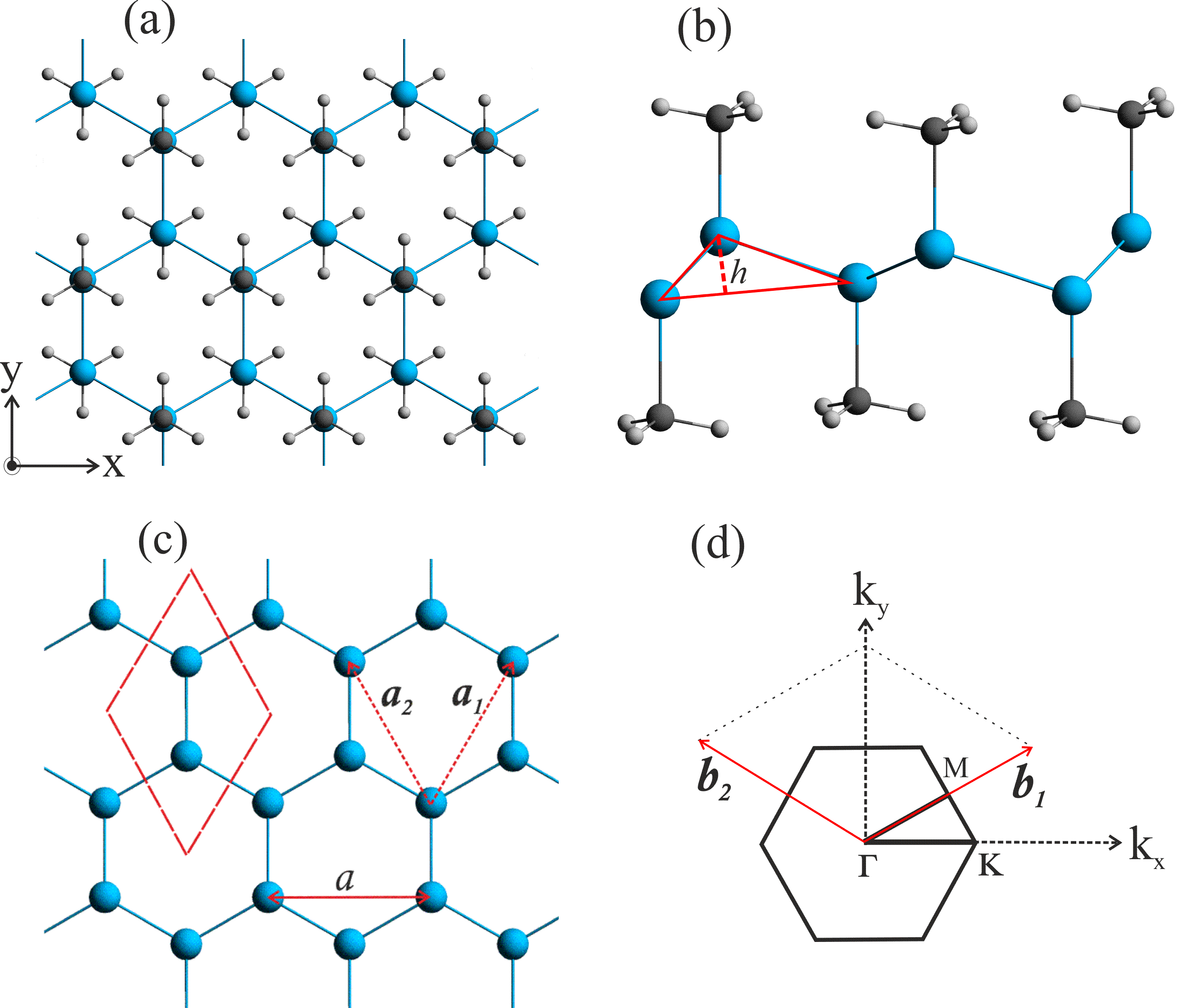}
\caption{ Schematic top (a) and side (b) views of  the monolayer GeCH$_3$ structure. Blue, 
black, and gray  balls indicate Ge, C, and H atoms, respectively. Ge atoms are sandwiched between two sheets of 
methyl groups. $h$ is the buckling of the structure. (c) Top view of the system eliminating 
the methyl group. ${\bm a_1}$ and ${\bm a_2}$ are the lattice vectors. (d) Fisrt Brillouin zone of the 
system with the reciprocal lattice vectors ${\bm b_1}$ and ${\bm b_2}$.}
\label{Lattice}
\end{figure}

The hexagonal atomic structure of  monolayer GeCH$_3$ and its geometrical parameters are 
shown in Figs.~\ref{Lattice}(a-c).
As shown in Figs.~\ref{Lattice}(a,b) it consists of three atomic layers where 
a buckled honeycomb sheet of Ge atoms is 
sandwiched between two outer methyl group layers. Each unit cell of  monolayer GeCH$_3$ 
consists of two Ge atoms and two CH$_3$ groups. Previous DFT calculations gave for the lattice constant $a=3.954$\AA , 
and the Ge-Ge and Ge-C bond lengths are 2.415\AA ~and 1.972\AA, respectively\cite{ma2014strain}. The buckling height, $h$, 
indicating the distance between two different Ge sublattices, is 0.788 \AA.

We have chosen the $x$ and $y$ axes along the armchair and zigzag directions, respectively. The $z$ axis is in the normal
direction to the plane of the  monolayer GeCH$_3$. With this definition of coordinates, the lattice vectors are written as
$\bm a_1=a/2(1,\sqrt{3}),~\bm a_2=a/2(-1,\sqrt{3})$,
 where the corresponding hexagonal Brillouin zone of the structure (see Fig.~\ref{Lattice}(d))
is determined by the reciprocal vectors
$\bm b_1=2\pi/a(1,\sqrt{3}/3),~\bm b_2=2\pi/a(-1,\sqrt{3}/{3})$.

\section{\label{ModelHamiltonian}Tight-Binding Model Hamiltonian}
Electronic structure of  monolayer GeCH$_3$ has been obtained by using DFT calculations in Ref.~\cite{ma2014strain}. It is shown that 
the low-energy electronic properties of this system are dominated by $s$, $p_x$ and $p_y$ atomic orbitals of Ge atoms. 
DFT calculations including SOC interaction have shown that applying an in-plane biaxial tensile strain induces a topological phase 
transition in the electronic properties of  monolayer GeCH$_3$~\cite{ma2014strain}. Although such a DFT approach, provides valuable 
information regarding the electronic properties of such system, it is limited to small computational unit cells. 
For example, large nanoribbons consisting of hundreds of atoms and including disorder require very large super-cells which go beyond present
 day computational DFT capability. This motivated us to derive a TB model for  monolayer GeCH$_3$ that is sufficiently accurate to describe 
the low-energy spectrum and the electronic properties of this system.

In the following we will propose a low-energy TB model Hamiltonian  that includes  SOC for  monolayer GeCH$_3$.
We show that our model is able to predict accurately the effect of strain on the electronic properties of the system.

\subsection{\label{NoSOC}Model Hamiltonian without SOC }
We propose a TB model including $s$ , $p_x$ and $p_y$ atomic orbitals with principal quantum number $n=4$ of Ge atoms to 
describe the low-energy spectrum of this system. The nearest-neighbor effective TB Hamiltonian without
SOC in the basis of $|s,p_x,p_y\rangle$ and in the second quantized representation is given by
\begin{equation}
H_0 =\sum_{i,\alpha}E_{i\alpha} c_{i\alpha}^\dagger c_{i\alpha}+ \sum_{\langle i,j\rangle,\alpha,\beta} t_{i\alpha,j\beta}
(c_{i\alpha}^{\dagger}c_{j\beta}+h.c),
\label{eqn:TBhamiltonian}
\end{equation} 
where $c_{i\alpha}^{\dagger}$ and $c_{i\alpha}$ represent the creation and annihilation 
operators for an electron in the $\alpha$-th orbital of the $i$-th atom, $E_{i\alpha}$ is the onsite energy of
$\alpha$-th orbital of the $i$-th atom and $t_{i\alpha,j\beta}$ is the nearest-neighbor hopping amplitude 
between $\alpha$-th orbital of $i$-th atom and $\beta$-th orbital  of $j$-th atom.
We will show that this effective model is sufficiently accurate to describe the low-energy spectrum of this system.

Note that the above Hamiltonian is quite different from the effective Hamiltonian that describes the electronic properties of
pristine germanene~\cite{kaloni2013stability}. In the pristine honeycomb structures of the group IV elements,
the effective low-energy spectrum  is
 described by the outer $p_z$ atomic orbitals. However, in monolayer GeCH$_3$, the $p_z$ orbitals mainly contribute
to the $\sigma$-bonding between Ge and C atoms to form the energy bands that are far from the Fermi level. Therefore, 
we will neglect the contribution of the $p_z$ orbitals of the Ge atoms and the other orbitals of the CH$_3$ molecule in our TB model.
\begin{figure}
\centering
\includegraphics[width=.45\textwidth]{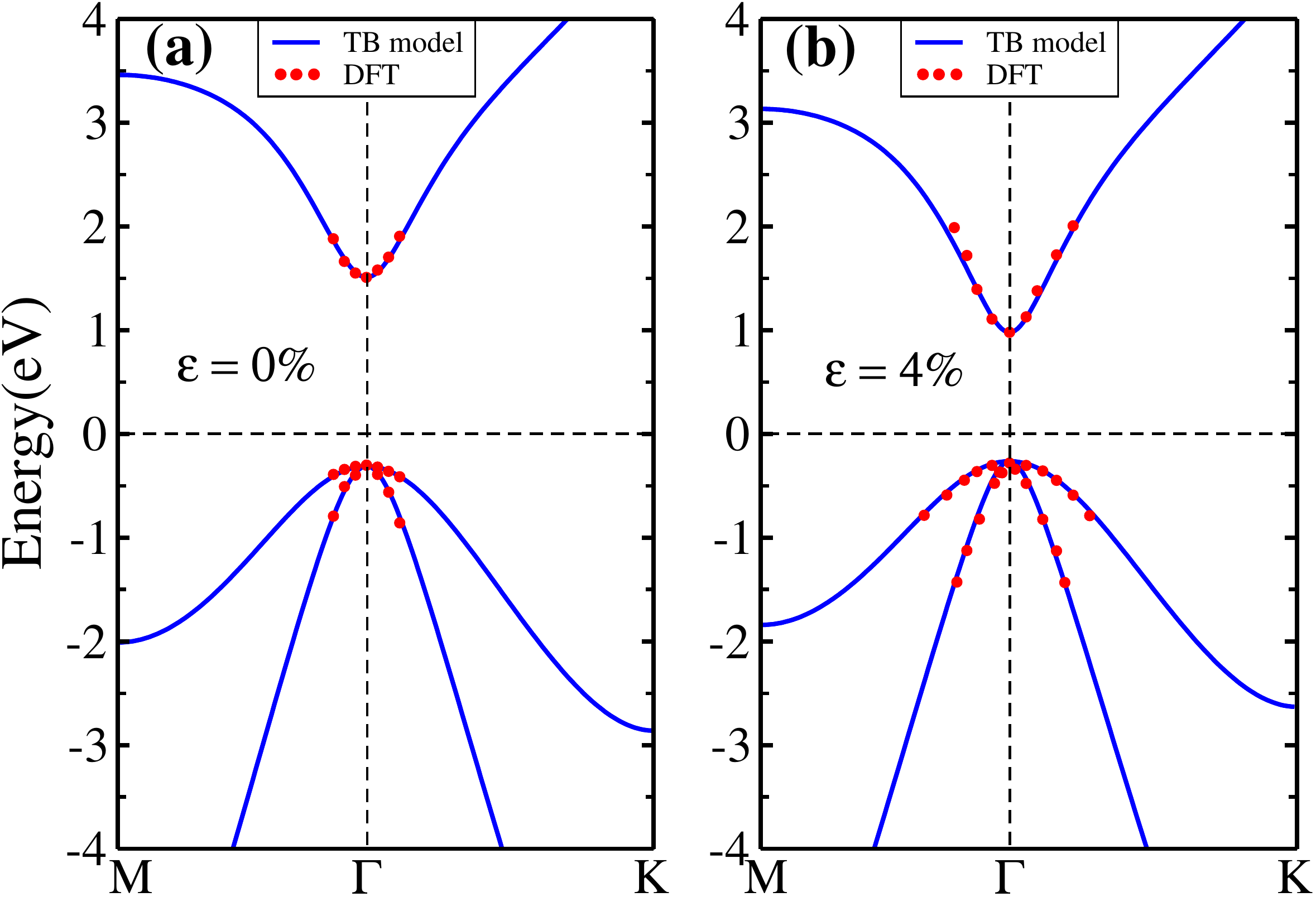}
\includegraphics[width=.45\textwidth]{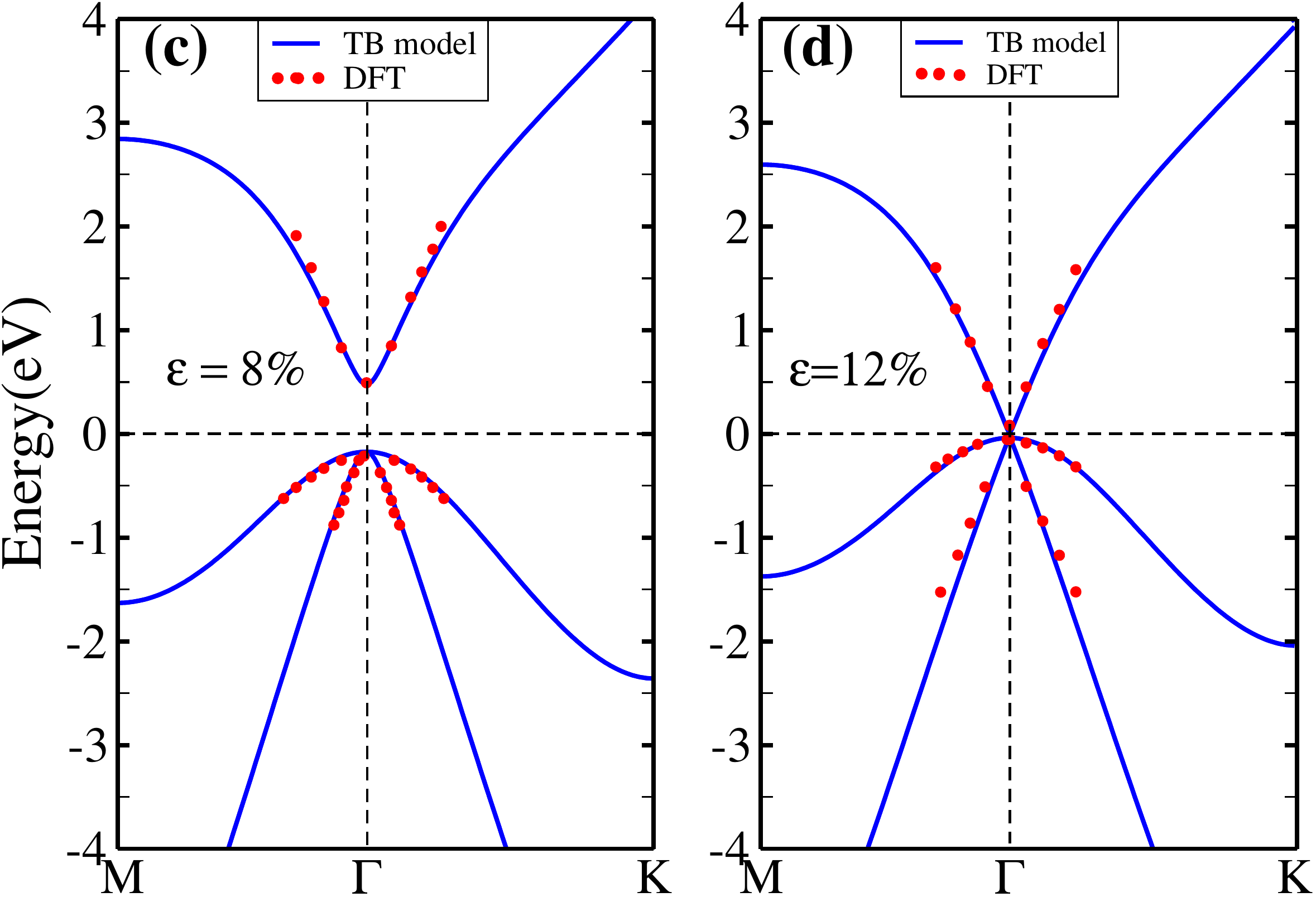}
\caption{The TB band structure of GeCH$_3$ isolated monolayer without SOC in the presence of (a) 0\%, (b) 4\%, (c) 8\%,
and (d) 12\% biaxial tensile strain. Symbols represent the HSE data taken from~\cite{ma2014sup}.}
\label{nosoc}
\end{figure}

\begin{table*}
\caption{The nearest neighbor hopping parameters between $s$ and $p$ orbitals are listed in the first
column. The second column represents the hopping integrals as a function of the standard Slater-Koster
parameters with direction dependent quantities. The third column are the nearest hopping parameters with the
inclusion of applied strain.}
\label{table1}
\begin{ruledtabular}
\begin{tabular}{ccc}
\textrm{Hopping parameters}&
\textrm{Without strain}&
\textrm{With biaxial strain}\\
\colrule
$t_{ss}$ & $V_{ss\sigma}$ & $t_{ss}^0[1-2\epsilon\cos\phi_0^2]$\\
$t_{sp_x}$ &$lV_{sp\sigma}$&$t_{sp_x}^0[1-2\epsilon\cos\phi_0^2+\eta\epsilon\tan\phi_0]$\\
$t_{sp_y}$ &$mV_{sp\sigma}$&$t_{sp_y}^0[1-2\epsilon\cos\phi_0^2+\eta\epsilon\tan\phi_0]$\\
$t_{p_xp_x}$ & $l^2V_{pp\sigma}+(1-l^2)V_{pp\pi}$&$t_{p_xp_x}^0[1-2\epsilon\cos\phi_0^2+2\eta\epsilon\tan\phi_0]-2\eta\epsilon\tan\phi_0V_{pp\pi}$\\
$t_{p_yp_y}$ &$m^2V_{pp\sigma}+(1-m^2)V_{pp\pi}$& $t_{p_yp_y}^0[1-2\epsilon\cos\phi_0^2+2\eta\epsilon\tan\phi_0]-2\eta\epsilon\tan\phi_0V_{pp\pi}$ \\
$t_{p_xp_y}$ & $lm(V_{pp\sigma}-V_{pp\pi)}$&$t_{p_xp_y}^0[1-2\epsilon\cos\phi_0^2+2\eta\epsilon\tan\phi_0]$ \\
\end{tabular}
\end{ruledtabular}
\end{table*}
With the above description, the hopping parameters of Eq.~(\ref{eqn:TBhamiltonian}) can be expressed in terms of the standard 
Slater-Koster parameters as listed in the middle column of Table~\ref{table1},
where $l=\cos\theta\cos\phi_0$ and $m=\sin\theta\cos\phi_0$ are, respectively, function of the cosine of
the angles between the bond connecting two neighboring 
atoms with respect to $x$ and $y$ axes.\\
Using the Fourier transform of Eq.~(\ref{eqn:TBhamiltonian}), and numerically diagonalizing the resulting Hamiltonian in $k$ space, 
 one can fit to the ab-initio results
in order to obtain the numerical values of the mentioned Slater-Koster parameters. The density functional calculation
results~\cite{ma2014sup}  including the  Heyd-Scuseria-Ernzerhof (HSE) 
functional approximation~\cite{heyd2003hybrid} are used to parametrize the TB model given by Eq.~(\ref{eqn:TBhamiltonian}). 
 We have listed the obtained numerical values
of these parameters in Table~\ref{table2}. The numerically calculated TB energy bands of  monolayer GeCH$_3$
in the absence of strain, as shown in Fig.~\ref{nosoc}(a), are 
in excellent agreement with the ab-initio results. The direct band gap of   monolayer GeCH$_3$ at the $\Gamma$ point is
1.82~eV.
\begin{table}
\caption{The values of the Slater-Koster parameters in units of eV as obtained from a fitting to the ab-initio results .}
\label{table2}
\begin{ruledtabular}
\begin{tabular}{cccccc}
\textrm{$V_{ss\sigma}$}&
\textrm{$V_{sp\sigma}$}&
\textrm{$V_{pp\sigma}$}&
\textrm{$V_{pp\pi}$}&
\textrm{$\epsilon_s$}&
\textrm{$\epsilon_p$}\\
\colrule
-2.20&2.62 &2.85&-0.85&-5.09&2.1
\end{tabular}
\end{ruledtabular}
\end{table}

\subsection{\label{strain}Strain effects}

 Applying strain to a system modifies its electronic properties \cite{bir1974symmetry}. 
This is due to the fact that it changes both the bond lengths and bond angles leading to a modulation 
of the hopping parameters that determine the electronic properties of the system.

An accurate prediction of the electronic properties of the system in the presence
of different types of strain, is a stringent test of the accuracy of our TB model. To this end, we now first calculate
the modification of the hopping parameters when biaxial tensile strain is applied to the plane of  monolayer GeCH$_3$.
Then we will study the modification of the energy spectrum  in the presence of such a strain to show that our results
agree very well with the DFT calculations. This particular type of strain
noticeably simplifies our calculations.
\begin{figure}[ht] 
\centering
\vspace{20pt}
\includegraphics[width=0.47\textwidth]{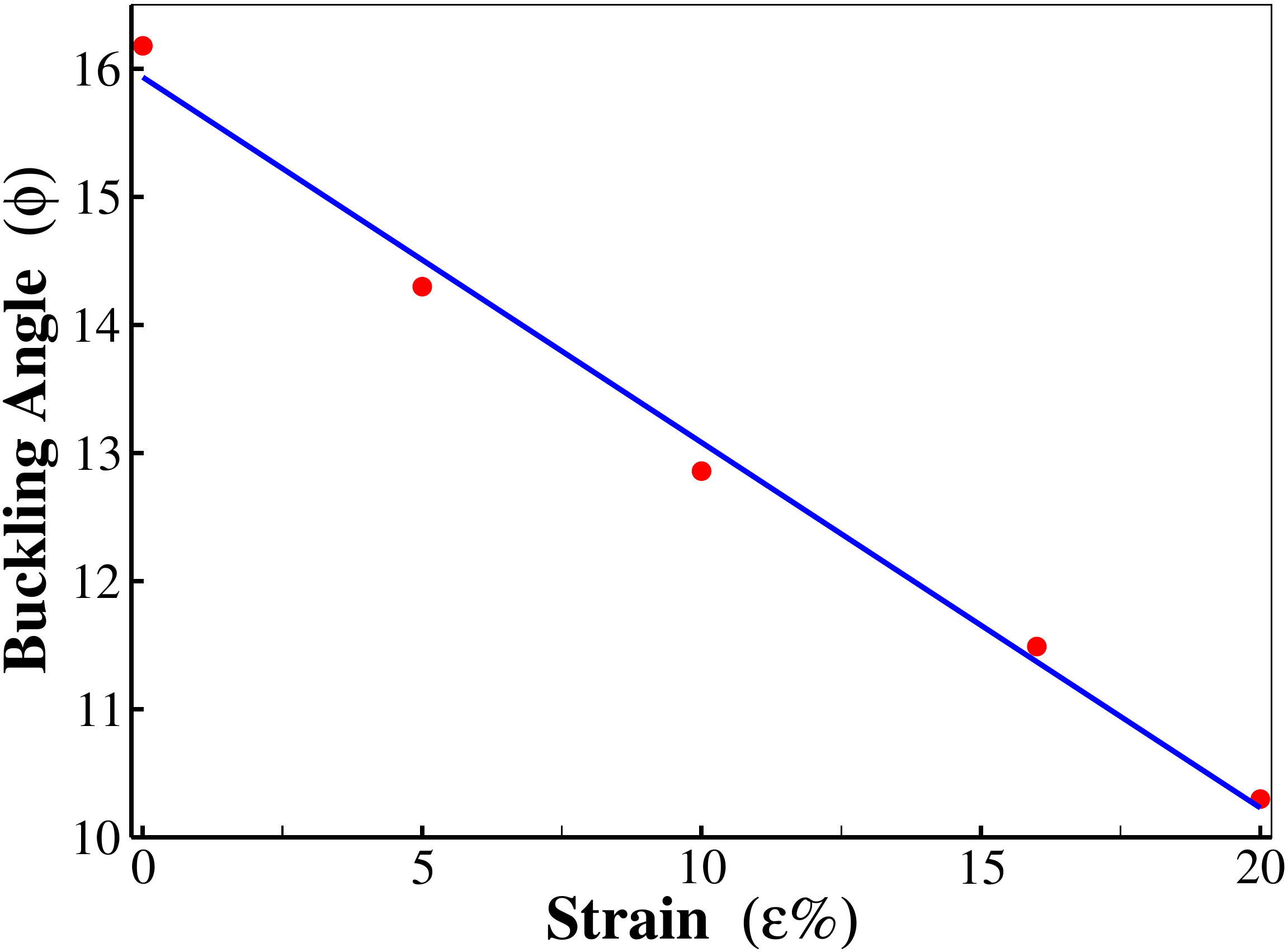}
\caption{Variation of the buckling angle as a function of biaxial tensile strain. Symbols represent the DFT data for germanene \cite{kaloni2013stability} and the solid line is the fit to this data.}
\label{buckling}
\end{figure}
 When biaxial tensile strain is applied in the plane of  monolayer GeCH$_3$ leaves the honeycomb nature of its lattice
intact and the initial lattice vectors ${\bm a^0_1}$ and ${\bm a^0_2}$ evolve to the deformed ones ${\bm a_1}$
and ${\bm a_2}$. Therefore, the vector $\bm r_0=(x_0,y_0,z_0)$, in 
the presence of in-plane strain is deformed into
$\bm r={(x,y,z)=[(1+\epsilon_x)x_0,(1+\epsilon_y)y_0,z_0]}$, where $\epsilon_x$ and $\epsilon_y$ are
the strain in the direction
of the ${ x}$ and ${y}$ axes, respectively.
In the following, for simplicity  we assume  that the strengths of the applied biaxial strains in
the two directions are equal, i.e.,
$\epsilon_x=\epsilon_y=\epsilon$. In the linear deformation regime, one can perform an expansion of the norm of
${r}$ to first order in $\epsilon_x$ an $\epsilon_y$ which results in
\begin{equation}
  r\simeq(1+\alpha_x\epsilon_x+\alpha_y\epsilon_y) r_0=[1+(\alpha_x+\alpha_y)\epsilon]r_0,
 \label{rstrain}
\end{equation}
where $\alpha_x={({x_0}/{r_0})}^2$ and $\alpha_y={({y_0}/{r_0})}^2$ are coefficients related to the geometrical structure
of GeCH$_3$. For the three nearest neighbor Ge atoms, one can write  $\alpha_x+\alpha_y=\cos^2\phi_0$, where $\phi_0$ is
the initial buckling angle. We note that in the presence of biaxial strain, the bond lengths and buckling angles are both altered.
 Thus, we consider their effects on the modification of the hopping parameters, simultaneously. Based on elasticity theory, 
we know that the main features of the mechanical properties in a covalent material are determined by the structure of the 
system and the strength of the covalent bonds. Therefore, one can expect that the change of the buckling angle in 
germanene \cite{kaloni2013stability} and GeCH$_3$ be akin. The variation of the buckling angle \cite{kaloni2013stability} as a function of
biaxial strain can be fit  to the linear form $\phi=\phi_0-\eta \epsilon$ (see Fig.~\ref{buckling}), where $\eta=-30$.

According to the Harrison rule \cite{harrison}, the standard Slater-Koster parameters related to $s$ and $p$ orbitals are proportional to
the bond length $r$ as $V_{\alpha\beta\gamma}\propto{1}/{r^2}$. Using Eq.~(\ref{rstrain}), the modified parameters 
are given  by 
\begin{equation}
 V_{\alpha\beta\gamma}=(1-2\epsilon\cos^2 \phi_0)V^0_{\alpha\beta\gamma}.
 \label{}
\end{equation}
One can then use the change of the buckling angle and the Slater-Koster parameters to obtain the modified hopping parameters as listed in the last 
column of Table~\ref{table1}, where $t_{\alpha\beta}^0$ represents the unstrained hopping parameters. For
instance, the new hopping parameter $t_{sp_{x}}$ can be approximated by 
 \begin{align}
  t_{sp_x}&=t^0_{sp_x}+\left(\frac{\partial t_{sp_x}}{\partial r}\right)_{r_0}\Delta r
  +\left(\frac{\partial t_{sp_x}}{\partial \phi}\right)_{\phi_0}\Delta\phi\nonumber\\
  &=t^0_{sp_x}-2\cos\theta\cos\phi_0V^0_{sp\sigma}
  \frac{\Delta r}{r_0}-\cos\theta\sin\phi_0V^0_{sp\sigma}\Delta\phi.
  \label{}
 \end{align}
Substituting ${\Delta  r}/{r_0}=\epsilon\cos^2\phi_0$  and  $\Delta\phi=-\eta\epsilon$  into
the above equation gives
\begin{equation}
 t_{sp_x}=t^0_{sp_x}[1-\epsilon(2\cos^2\phi_0-\eta\tan\phi_0)].
 \label{}
\end{equation}
In a similar way, one can obtain the other modified  hopping parameters in order to study the evolution of the energy spectrum of
 monolayer GeCH$_3$ as a function of applied biaxial tensile strain.

 Straightforward substitution of the new hopping parameters in 
 Hamiltonian, Eq.~(\ref{eqn:TBhamiltonian}), gives the Hamiltonian for the strained system. The calculated TB energy spectrum in the 
 presence of biaxial tensile strain with strengths of $4\%$, $8\%$, and $12\%$ are shown in 
 Figs.~\ref{nosoc}(b), (c) and (d), which are in excellent agreement with the DFT results \cite{ma2014strain,ma2014sup}.
 \begin{figure}
\centering
\includegraphics[width=.47\textwidth]{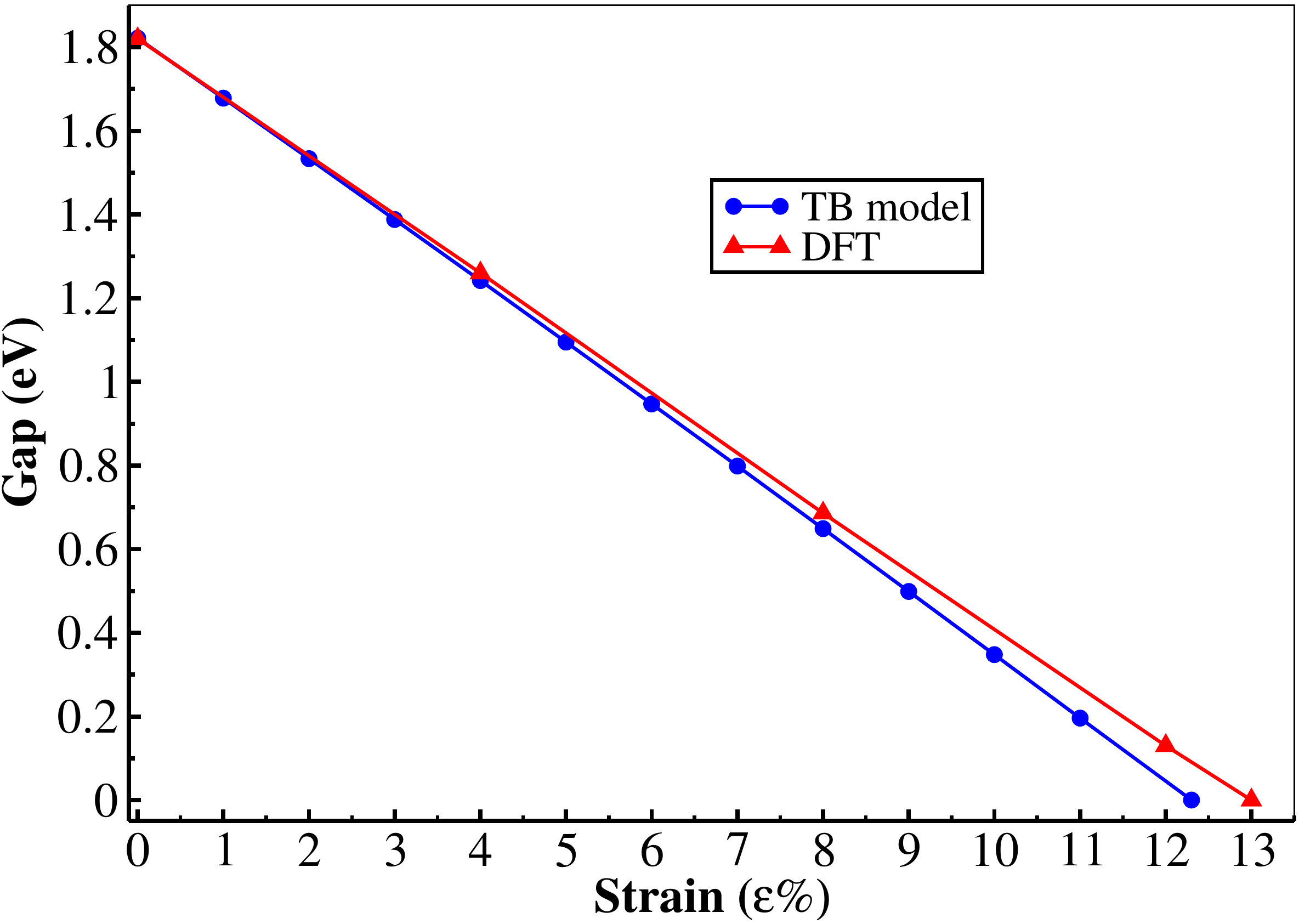}
\caption{Comparison of the variation of energy band gap vs. biaxial strain between TB model and HSE 
calculations~\cite{ma2014strain}.}
\label{Gap}
\end{figure}
We show in Fig.~\ref{Gap} the dependence of the band gap of GeCH$_3$ as function of biaxial tensile strain. Notice the 
good agreement between both DFT and TB approaches demonstrating the validity of our proposed TB model.

\subsection{\label{SOC_Hamiltonian}Spin-Orbit coupling}
Spin-orbit interaction is a relativistic correction to the Schr\"{o}dinger equation. It can significantly affect the electronic 
properties of systems that consists of heavier elements. In such systems, the major part of SOC  originates from the 
orbital motion of electrons close to the atomic nuclei. In the Slater-Koster approximation, one can assume an effective spherical
atomic potential $V_i({\bm r})$, at least in the region near the nucleus. Therefore, one can 
substitute $\nabla V_i({\bm r})=({dV_i}/{dr}){\bm r}/{r}$ and ${\bm s}={\hbar}/{2}{\bm \sigma}$ 
into the general form for the SOC term~\cite{li2015giant,zhao2015driving}
\begin{equation}
H_{SOC}=-\frac{\hbar}{4m_0^2c^2}(\nabla V\times {\bm p})\cdot{\pmb \sigma},
\end{equation}
to obtain the SOC in the form of  
\begin{equation}
H_{SOC}=\lambda(r){\bm L}\cdot\pmb{\sigma},
\end{equation}
where $\lambda(r)={1}/{2m_0^2c^2r}(dV/dr)$ is a radial function whose value depends on the 
type of 
atomic species. In the above equations, $\hbar, m_0, c$ and ${\bm p}$, are Plank constant, free mass of electron, speed of light, and
momentum, respectively; and ${\pmb \sigma},{\bm L}$ and ${\bm s}$ represent the Pauli matrices, angular momentum operator and
electron spin operator, respectively.

Using the well known ladder  operators $ L_\pm$ and $S_\pm$, one can obtain the matrix representation of the SOC
Hamiltonian in the basis set of $|s_{1},p_{x1},p_{y1},s_{2},p_{x2},p_{y2}\rangle \otimes |\uparrow,\downarrow\rangle$
for monolayer GeCH$_3$ with matrix elements 
\begin{equation}
\langle\alpha_i|H_{SOC}|\beta_i\rangle=\lambda_i<{\bm L}\cdot{\pmb \sigma}>_{\alpha\beta},
\label{hsoc2}
\end{equation}
where $\alpha_i$ and $\beta_i$ represent the atomic orbitals of $i$-th atom. Note that since the two atom basis in the unit
cell of the  monolayer GeCH$_3$ are the same, we have $\lambda_1=\lambda_2=\lambda$.

Thus, the representation of the SOC Hamiltonian in the above mentioned basis is 
\begin{equation}
H_{SOC}=
\begin{bmatrix}
H_{SOC}^{\uparrow\uparrow}&H_{SOC}^{\uparrow\downarrow}\\H_{SOC}^{\downarrow\uparrow}&H_{SOC}^{\downarrow\downarrow}
\end{bmatrix},
\end{equation}
whose elements are $6\times6$ matrices with $H_{SOC}^{\uparrow\downarrow}=H_{SOC}^{\downarrow\uparrow}={\bm 0}$, and

\begin{equation}
  H_{SOC}^{\uparrow\uparrow}=\lambda
  \begin{bmatrix}
  0&0&0\\0&0&-i\sigma_z\\0&i\sigma_z&0\
  \end{bmatrix}
  ,~~H_{SOC}^{\downarrow\downarrow}=\lambda
  \begin{bmatrix}
   0&0&0\\0&0&i\sigma_z\\0&-i\sigma_z&0\  
  \end{bmatrix}.
\end{equation}

\begin{figure}[ht] 
\centering
\vspace{20pt}
\includegraphics[width=0.3\textwidth]{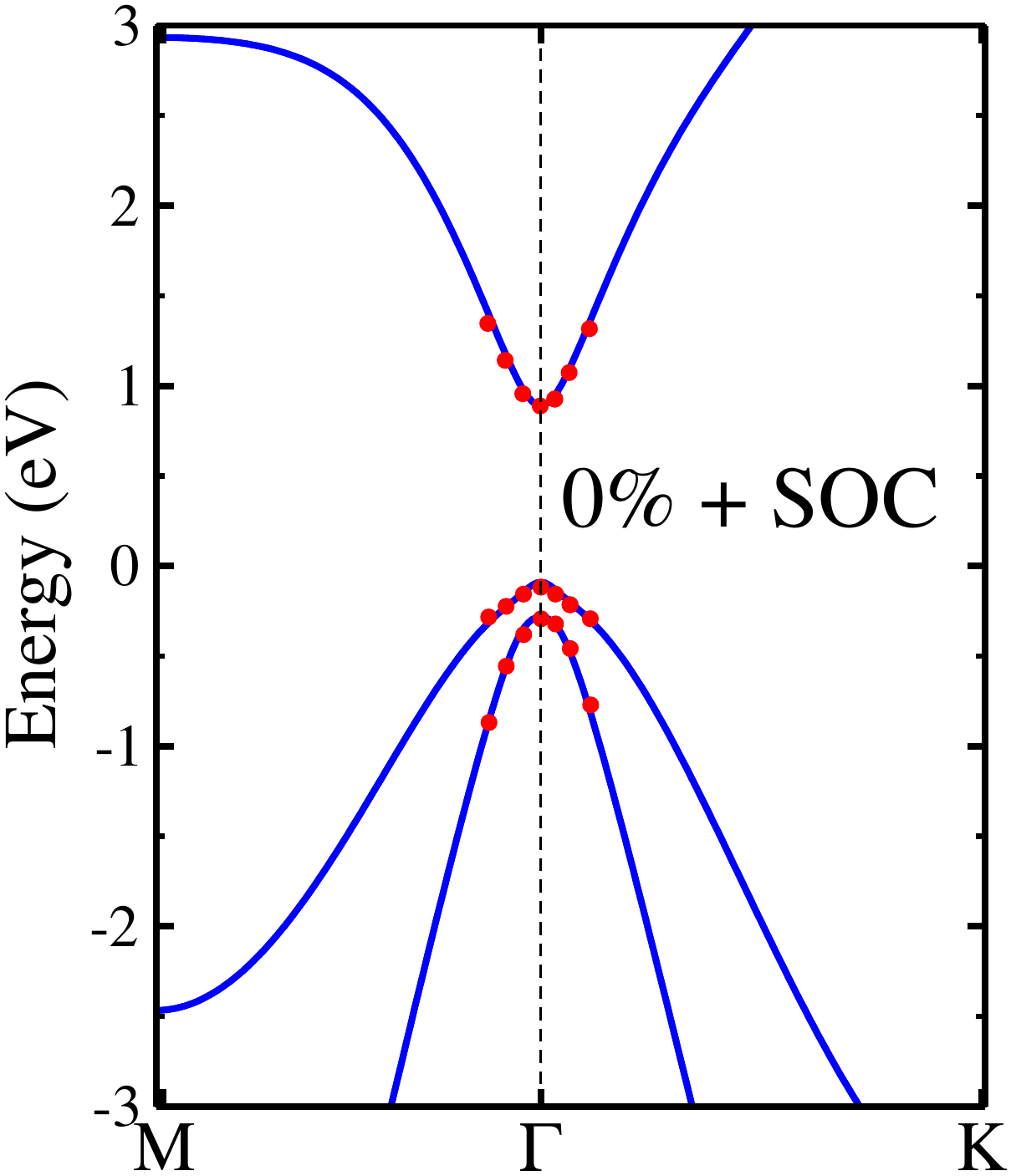}
\caption{The multi-orbital TB spectrum of GeCH$_3$ monolayer with SOC. Symbols represent the LDA data taken from~\cite{ma2014strain}.}
\label{LDA}
\end{figure}

The value of the strength $\lambda$ of the SOC should be chosen either in agreement with experiment or by fitting
the TB bands to the ab-initio results near some $k$ points such that it gives the correct band gap.
 In order to evaluate the strength of the SOC for Ge atoms in  monolayer GeCH$_3$, we fitted the 
 spectrum obtained from our multi-orbital TB model
to the one from density functional calculations within the local density approximation (LDA) for the 
exchange correlation in Ref.~\cite{ma2014strain}. 
 As shown in Fig.~\ref{LDA}, there is excellent agreement between the TB spectrum and the DFT results for the 
SOC strength $\lambda=0.096$~eV. We adopt this SOC strength in the following calculations of the TB spectrum when 
we use the hopping parameters from Table~\ref{table2}. 

\begin{figure}[ht] 
\centering
\vspace{20pt}
\includegraphics[width=0.45\textwidth]{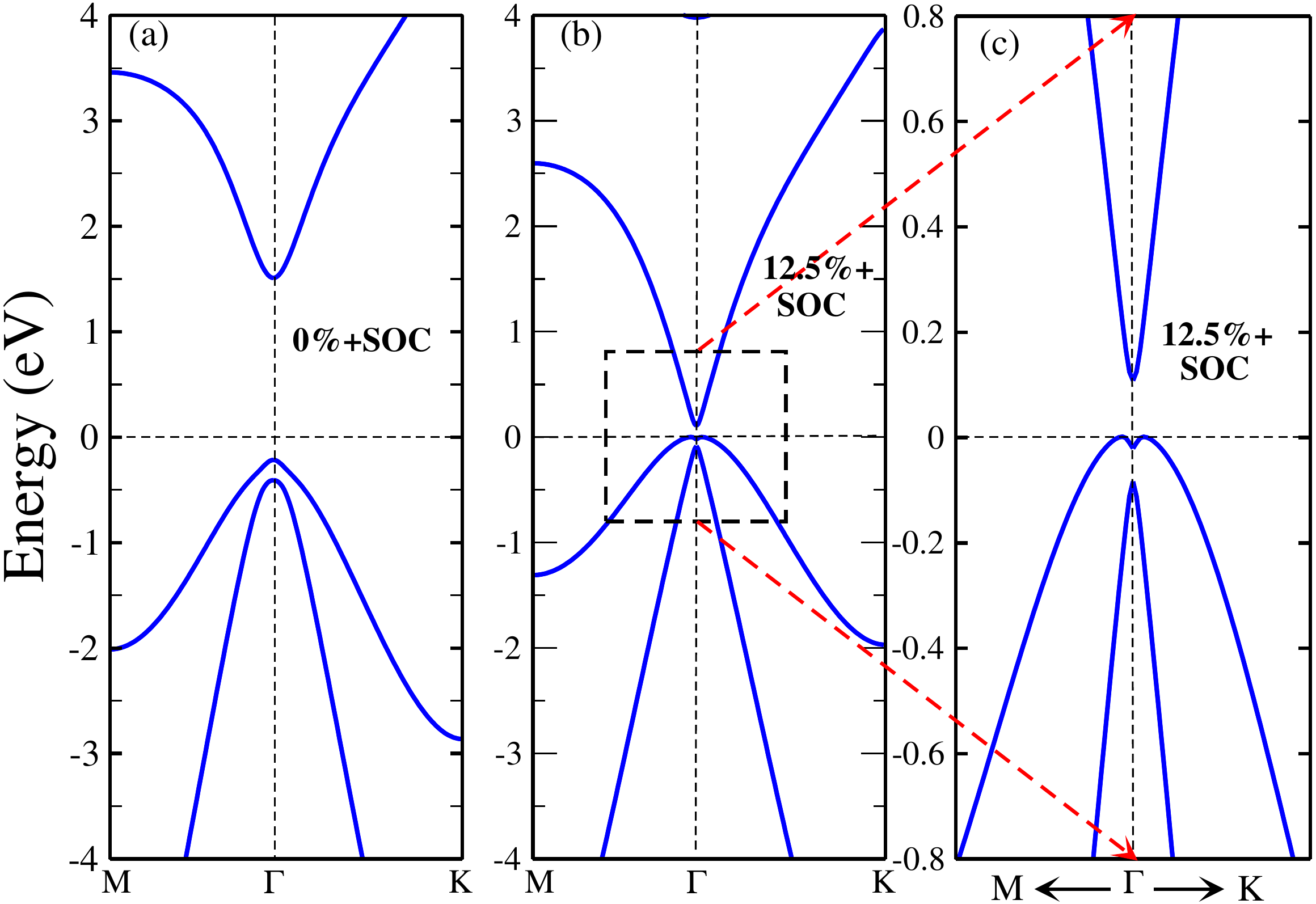}
\caption{The TB band structure of GeCH$_3$ isolated monolayer with SOC in the presence of (a) 0\%, and (b) 12.5\% biaxial tensile strain.
(c) Zoomed-in view of (b).}
\label{with_SOC}
\end{figure}

The TB energy spectrum
of  monolayer GeCH$_3$ are shown
in Figs.~\ref{with_SOC} (a) and (b) for 0\% and 12.5\% strain, respectively. 
\begin{figure}[ht] 
\centering
\vspace{20pt}
\includegraphics[width=0.45\textwidth]{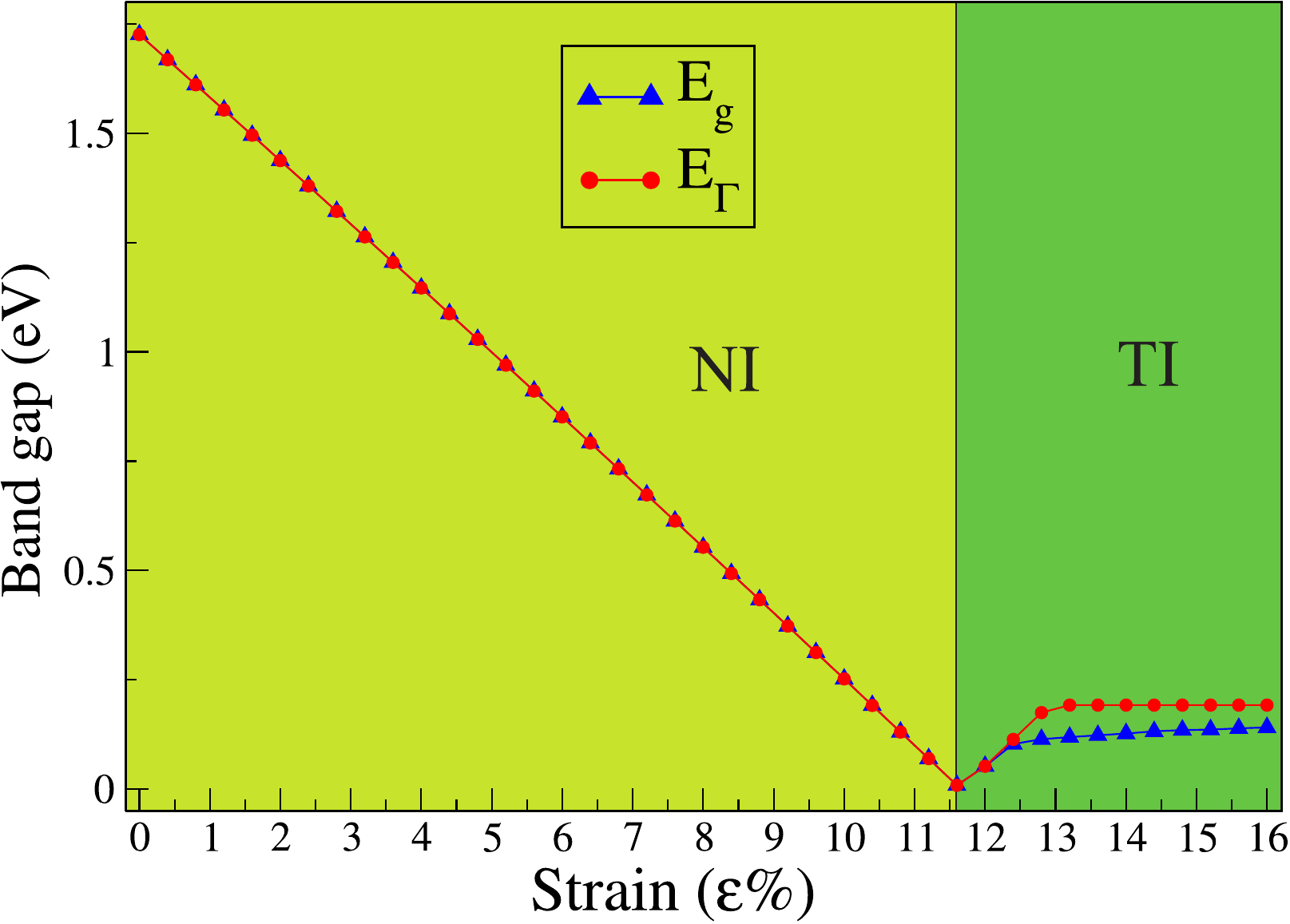}
\caption{The calculated band gaps of  monolayer GeCH$_3$ as a function of biaxial strain at the $\Gamma$ point, $E_\Gamma$,
and the global gap $E_g$. The two distinct colored regions show the different trivial and band inverted phases.}
\label{gap_soc}
\end{figure}
Note that due to the presence of time reversal and inversion symmetry, each band in the energy spectrum of 
 monolayer GeCH$_3$ is doubly degenerate. 
 As shown in Fig.~\ref{gap_soc},
by applying biaxial tensile strain, the global band gap located at $\Gamma$ gradually
decreases and eventually a band inversion occurs at 11.6\% strain.
By further increasing strain,
the induced band gap due to SOC, (see Figs.~\ref{with_SOC}(b), and (c)) becomes indirect, and at a 
reasonable strength of 12.8\% reaches the value of 115 meV.

 One can use the TB spectrum of Figs.~\ref{with_SOC}, to calculate the effective masses of electrons 
and holes near the conduction band minimum (CBM) and the valence band maximum (VBM). The results, in unit of 
free electron mass $m_0$, are listed in Table~\ref{table3}
for 0\%, 6\%, 9\%, and 12.5\% biaxial tensile strain.
Note that, the electron and hole effective masses near the CBM and VBM 
along the two directions of $\Gamma$-K and $\Gamma$-M are the same.
\begin{table}
\caption{The effective mass of electron
and hole near the CBM and VBM in unit of 
free electron mass $m_0$. The electron and hole effective masses along the two directions of $\Gamma$-K and $\Gamma$-M are
the same.}
\label{table3}
\begin{ruledtabular}
\begin{tabular}{ccc}
\textrm{ ~~~~~~~~~~~~~~Strain ($\epsilon$)\textbackslash Effective mass ($m/m_0$)}&
\textrm{Electron}&
\textrm{Hole}\\
\colrule
~0\% &0.135 &0.157\\
~~~~6\% &0.074 &0.105\\
~~~~9\% &0.045 &0.058\\
~~~~12.5\% &0.033 &0.316\\

\end{tabular}
\end{ruledtabular}
\end{table}

Another way to test the validity of our TB model, is its ability to predict a possible topological phase transition in the
electronic properties of monolayer GeCH$_3$. In the next section we will study the strain-induced topological phase in
   monolayer GeCH$_3$ using our TB model.
 
 \section{\label{Numericalresults} topological phase transition of
 monolayer G\lowercase{e}CH$_3$ under strain}

In the previous section, using the TB model including
SOC, we showed that monolayer  GeCH$_3$ is a NI. We also showed
that one can manipulate its electronic properties by applying in-plane biaxial strain.
It is clear from Eq.~(\ref{hsoc2}) that SOC preserves the TRS.
Thus, the  monolayer GeCH$_3$ can exhibit a QSH phase when its energy 
spectrum is manipulated by an external parameter that does not break TRS.
The $\mathbb{Z}_2$ classification is a well known approach to distinguish between
the two different NI  and TI phases~\cite{kane,hasan}.
In the following, we briefly introduce the lattice version of the Fu-Kane formula~\cite{fu1},
to calculate the $\mathbb{Z}_2$ invariant. Then, we show numerically that by applying biaxial tensile
strain, a change in the bulk topology of monolayer GeCH$_3$ occurs.
\subsection{\label{z2}Calculation of the $\mathbb{Z}_2$ invariant }
The Fu-Kane formula~\cite{fu1}, for the calculation of the $\mathbb{Z}_2$ invariant is given by
\begin{equation}
\text{\footnotesize $\mathbb{Z}_2=\frac{1}{2\pi i}\left [\oint_{{\partial\textrm{HBZ}}}d\bm{k}\bm{\cdot\mathcal{A}}(\bm{k})-
\int_{{\textrm{HBZ}}}d^2k \mathcal{F}(\bm{k})]\right]\textrm {(mod 2)}$},
\label{z2-1}
\end{equation}
where the integral is taken over half the Brillouin zone as denoted by $\small{\textrm{HBZ}}$.
Here, the Berry gauge potential $\bm{\mathcal{A}}(\bm{k})$, and the Berry field 
strength $\mathcal{F}(\bm{k})$ are 
given by $\sum_n\langle u_n(\bm{k})|\nabla_n u_n(\bm{k})\rangle$, and 
$\nabla_{\bm k} \times {\mathcal{A}(\bm{k})\mid_z}$, respectively; where $u_n(\bm{k})$ represents
the periodic part of the Bloch wave function with band index $n$, and the summation in
$\bm{\mathcal{A}}(\bm{k})$ runs over all occupied states.\\
 Note that, in this approach one has to do some gauge fixing procedure~\cite{fukui1}
 to fulfill the TRS constraints and the periodicity of the
 $k$ points  which are related by a reciprocal lattice vector $\bm G$. 
 Moreover, due to the TRS and the inversion symmetry in  monolayer GeCH$_3$, each band
 is at least doubly degenerate. Therefore, one needs to generalize the definition of 
 $\bm{\mathcal{A}}$ and $\mathcal{F}$ to non-Abelian gauge field analogies~\cite{hatsugai2004}
constructed from the 2M dimensional ground state multiplet $|\psi(k)\rangle=(|u_1(k)\rangle,...,|u_{2M}(k)\rangle)$, 
associated to the Hamiltonian $\mathcal{H}(k)|u_n(k)\rangle= E_n(k)|u_n(k)\rangle $~\cite{fukui1,hatsugai2004}.\\
\begin{figure}[ht] 
\centering
\vspace{20pt}
\includegraphics[width=0.48\textwidth]{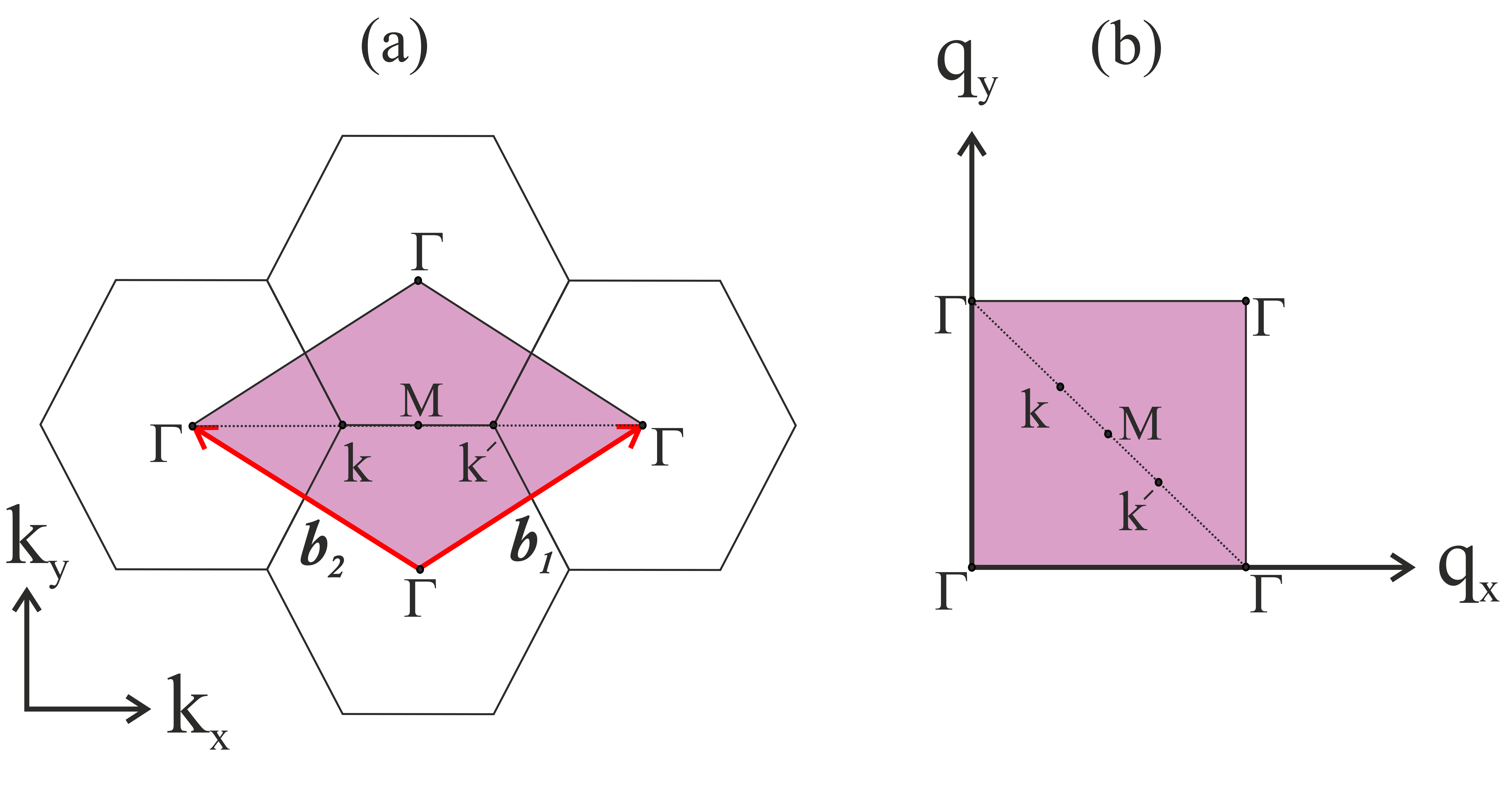}
\caption{Conversion of the equivalent (a) rhombus 
shape of the honeycomb Brillouin zone in $k$ space into a (b) unit square in $q$ space.
}
\label{Bz}
\end{figure}

\begin{figure}[ht] 
\centering
\vspace{20pt}
\includegraphics[width=0.48\textwidth]{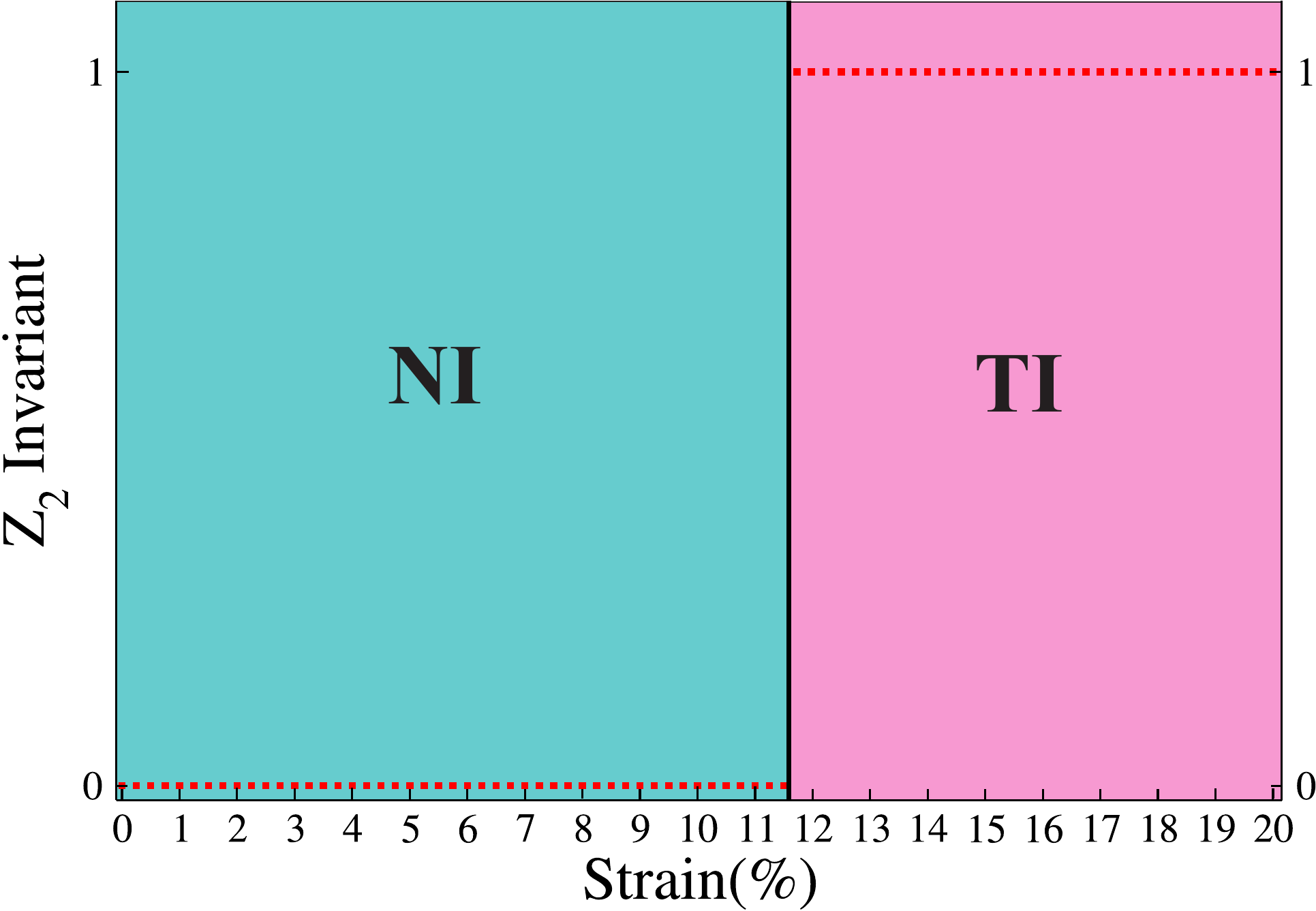}
\caption{Calculation of $\mathbb{Z}_2$ invariant for  monolayer GeCH$_3$ in the 
presence of biaxial strain. The two NI and TI phases are represented by regions of different colors and delimited
by a black line at the critical value of $11.6\%$.
}
\label{z2}
\end{figure}
In order to compute the $\mathbb{Z}_2$ invariant, a lattice version of Eq.~(\ref{z2-1}) is more favorable for 
numerical calculations. To this end, one can simply convert the equivalent rhombus 
shape of the honeycomb Brillouin zone in $k$ space  as shown in Figs.~\ref{Bz}(a) and (b),
into a unit square in $q$ space by the following change of variables

\begin{equation}
k_x=\frac{2\pi}{a}(q_x-q_y), ~~~~ k_y=\frac{2\pi}{\sqrt{3}a}(q_x+q_y).
\end{equation}

This, allows us to use the more simple lattice version of Eq.~(\ref{z2-1})~\cite{fukui1}
  \begin{equation}
\small\mathbb{Z}_2=\frac{1}{2 \pi i}\left[ \sum_{{\textit q_l}\in \small{\partial\textrm{HBZ}}}A_x({\textit q_l}) - 
\sum_{{\textit q_l}\in \small\textrm{HBZ}} F_{xy} ({\textit q_l})\right]\textrm {(mod 2)},
\label{z2-2}
\end{equation}
where the lattice sites of the Brillouin zone are labeled by $q_l$. Thus the above mentioned gauge 
fixing procedure and TRS constraints are applied on the equivalent $q$ points. Using the 
so-called unimodular link variable~\cite{fukui1}
\begin{equation}
U_{\hat\mu}({\textit q_l})=\frac{\textrm {det}  \psi^\dagger(\textit q_l)\psi(\textit q_l+ \hat{\mu})}{|\textrm {det}  \psi^\dagger(\textit q_l)\psi(\textit q_l+ \hat{\mu})|},
\end{equation}
where $\hat{\mu}$ denotes a unit vector in the $q_x$-$q_y$ plane, one can define the Berry potential and Berry field 
in Eq.~(\ref{z2-2}) as
\begin{eqnarray}
 A_{x}({\textit q_l})&=&\ln U_x({\textit q_l}), \\
\small F_{xy}(\textit q_l)&=&\ln\frac{U_x(\textit q_l)U_y(\textit q_l+\hat{x})}{ U_y(\textit q_l)U_x(\textit q_l+\hat{y})}.
\label{FFF}
\end{eqnarray}
Note that both the Berry potential and the  Berry field strength are defined within the
branch of $A_{x}({\textit q_l})/i\in(-\pi,\pi)$ and $F_{xy}(\textit q_l)/i\in(-\pi,\pi)$.\\

The numerical results of the  $\mathbb{Z}_2$ invariant are shown in Fig.~\ref{z2}.
As seen, for $\epsilon<11.6$\%,  monolayer GeCH$_3$ is a NI and at the critical value of 
$\epsilon=11.6$\%, the $\mathbb{Z}_2$ invariant jumps from 0 to 1, indicating a strain-induced TI phase transition in
 the electronic properties of the system. The topologically  
protected global bulk gap for a strain of 12.8\% is 115~meV, which is much larger than the thermal
energy at room temperature and therefore the monolayer GeCH$_3$  is an excellent candidate for strain related applications.\\
 In the next subsection we examine the formation of topologically protected edge states
 in a typical nanoribbon with zigzag edges when the system is driven into the TI phase by applying
biaxial tensile strain.

\subsection{\label{SOC_Hamiltonian}Electronic properties of GeCH$_3$ nanoribbons under strain}
The appearance of helical gapless states at the edge of a 2D topological insulator, is a crucial consequence of its nontrivial
bulk topology. In the previous section, we showed that a jump from 0 to 1 in the $\mathbb{Z}_2$ invariant for biaxial strain at
$\epsilon>11.6\%$ takes place, demonstrating a topological phase transition in the electronic properties of  monolayer GeCH$_3$.
 As an example, in this subsection, we study the 1D energy bands of GeCH$_3$ nanoribbons with zigzag edges in the presence of biaxial
 tensile strain. Our TB model predicts the appearance of topologically protected edge states with increasing strain when the  $\mathbb{Z}_2$
 invariant becomes 1. We denote the width of the zigzag GeCH$_3$ nanoribbon (z-GeCH$_3$-NR)  by N, which is the number of zigzag chains
 across the ribbon width. To calculate the energy spectrum of a z-GeCH$_3$-NR with width N, we construct its 
 supercell Hamiltonian ($H^{SC}$) in the basis of 
$|\psi\rangle\equiv|s_{H_0},s_{1},p_{x1},p_{y1},...,s_{2N},p_{x2N},p_{y2N},s_{H_1}\rangle \otimes |\uparrow,\downarrow\rangle$ 
where $s_i$, $p_{xi}$, and $p_{yi}$ represent the $s$, $p_{x}$, and $p_{y}$ orbitals of Ge atoms along the nanoribbon width.
 \begin{figure}[ht] 
\centering
\vspace{20pt}
\includegraphics[width=0.48\textwidth]{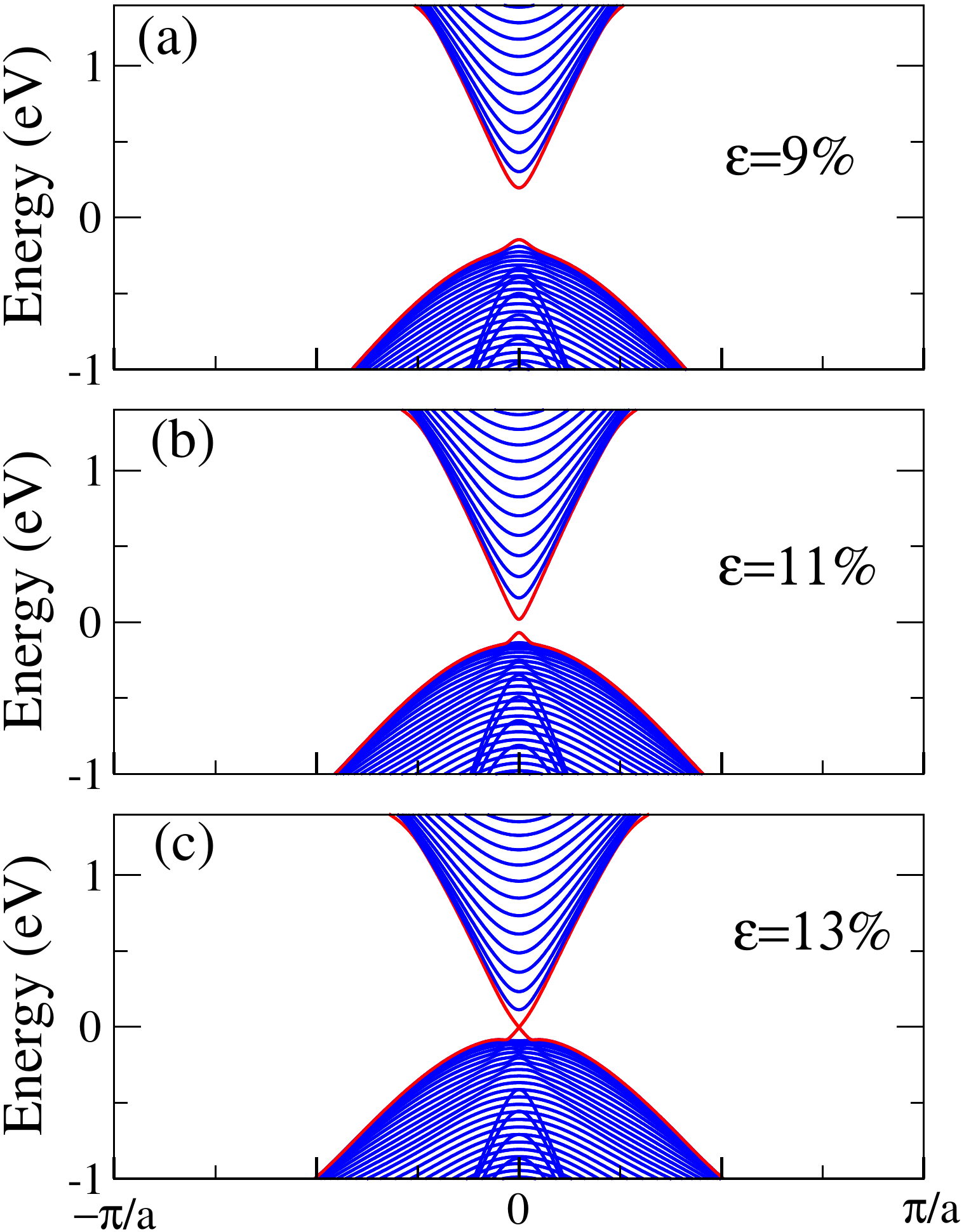}
\caption{The 1D energy bands of z-GeCH$_3$-NR for $N=40$ in the presence of		 
(a) 9\%, (b) 11\%, and (c) 13\% biaxial tensile strain.
}
\label{band_zig}
\end{figure}
$|s_{H_0}\rangle$ and $|s_{H_1}\rangle$ represent the atomic orbitals of H atoms that are introduced to passivate the Ge 
 atoms on each edge, respectively. We assume that the width of the nanoribbon is large enough that the interaction between the 
 two edges is negligible, and one can safely neglect the tiny change of the hopping parameters due to the passivation procedure. 
 Therefore, one can write the matrix elements of the nanoribbon Hamiltonian $H^{SC}=H_0^{SC}+H_{SOC}^{SC}$ as 
 \begin{align}
 \label{14}
M_{i\alpha,j\beta}^{\sigma\sigma^\prime}&=\langle\psi|H^{SC}|\psi\rangle_{i\alpha,j\beta}^{\sigma\sigma^\prime}\nonumber\\
&=E_{i\alpha}\delta_{ij}\delta_{\alpha\beta}\delta_{\sigma\sigma^\prime}\nonumber\\&
+\delta_{\sigma\sigma^\prime}\sum_nt_{i\alpha,j\beta}
e^{i\bm k\cdot \bm R_{0n}}+\lambda_i\delta_{ij}<\bm L\cdot\pmb\sigma>_{\alpha\beta}^{\sigma\sigma^\prime},
 \end{align}
 where $i,j$ are the basis site indices in a supercell; $\alpha,\beta$ denote the atomic orbitals; $\sigma, \sigma^\prime$ denote the
 spin degrees of freedom; and ${\bm R_{0n}}$ is the translational vector of the $n$-th supercell. 
 The corresponding onsite energy of Ge atoms and the hopping parameters pertinent to the Ge-Ge bonds are substituted from
 Tables~\ref{table1} and ~\ref{table2}. Moreover, one has to define the onsite energy $E_H^s$, and the hopping parameters $t_{H,Ge}^{ss}$ and 
 $t_{H,Ge}^{sp_{y}}$  in the above
 equation corresponding to the matrix elements related to the H-Ge bond. We adopt from the 
 fitting procedure the numerical values $E_H^s=-2.54$ eV, 
 $t_{H,Ge}^{ss}=V_{H,Ge}^{ss}=-4.54$ eV, and $t_{H,Ge}^{sp_y}=\pm V_{H,Ge}^{sp}$ with $V_{H,Ge}^{sp}=0.5$ eV where +(-) denotes the lower 
 (upper) H-Ge edge bonds.
 One can diagonalize the corresponding TB Hamiltonian, Eq.~(\ref{14}), in order to obtain the energy spectrum.
 By applying  biaxial tensile strain we found that the band gap of the nanoribbon
 gradually decreases and eventually the metallic edge states protected by TRS
 appear for a strain value where a band inversion takes place in the TB energy spectrum of  bulk  monolayer GeCH$_3$.
 The numerically calculated 
 energy bands of z-GeCH$_3$-NR  with $N=40$ in the presence of 9\%, 11\%, and 13\% biaxial tensile strain
  are shown in Figs.~\ref{band_zig}(a), (b), and (c), respectively.
 This demonstrates a topological phase transition from the NI to the QSH phase in the electronic properties
 of  monolayer GeCH$_3$.

\section{\label{Conclusion}Conclusions}
To conclude, we have proposed an effective TB model with and without SOC for  monolayer GeCH$_3$ 
including $s$, $p_x$, and $p_y$ orbitals per atomic site.
Our model reproduces the low-energy spectrum of   monolayer GeCH$_3$ in 
excellent agreement with ab-initio results. It also predicts accurately the evolution of the band gap 
in the presence of  biaxial tensile strain. By including the SOC, this band gap manipulation
leads to a band inversion in the electronic properties
of   monolayer GeCH$_3$, giving rise to a topological phase transition from NI to QSH.
Our model predicts that this phase transition takes place for 11.6\%  biaxial tensile strain
as verified by the $\mathbb{Z}_2$ formalism.
The topologically protected global bulk gap at a strain of 12.8\% is 115~meV,
which is much larger than the thermal
energy at room temperature and makes monolayer  GeCH$_3$ a promising candidate for future
applications. We also showed the emergence of topologically protected edge states in 
a typical z-GeCH$_3$-NR in the presence of biaxial strain larger than 11.6\%. This is an additional confirmation of the existence of the TI phase 
in the electronic properties of monolayer GeCH$_3$. 

\nocite{*}

\bibliography{GeCH3}

\end{document}